\documentclass[aps,prb,twocolumn,superscriptaddress,showpacs,floatfix,10pt]{revtex4-1}
\usepackage{amsmath}
\usepackage{amsfonts}
\usepackage{graphicx}
\usepackage{color}
\usepackage{epstopdf}
\usepackage{natbib}
\usepackage{xfrac}
\usepackage[normalem]{ulem}

\newcommand{\kk}{\mathbf{k}}
\newcommand{\rr}{\mathbf{r}}

\newcommand{\AAA}{\mathbf{A}}
\newcommand{\EE}{\mathbf{E}}
\newcommand{\BB}{\mathbf{B}}
\newcommand{\En}{\mathcal{E}}
\newcommand{\JJ}{\mathbf{J}}

\begin{document}
\title{Boltzmann approach to high-order transport: the non-linear and non-local responses}

\author{M. Battiato}
\affiliation{Institute of Solid State Physics, Vienna University of Technology,  Vienna, Austria}
\email[]{marco.battiato@ifp.tuwien.ac.at}
\author{V. Zlatic}
\affiliation{Institute of Physics, Zagreb, Croatia}
\author{K. Held}
\affiliation{Institute of Solid State Physics, Vienna University of Technology,  Vienna, Austria}

\date{\today}

\begin{abstract}

The phenomenological textbook equations for the charge and heat transport are extensively used in a number of fields ranging from semiconductor devices to thermoelectricity. 
We provide a rigorous derivation of transport equations by solving the Boltzmann equation in the relaxation time approximation 
and show that  the currents can be rigorously represented by an expansion in terms of the 'driving forces'. 
Besides the linear and non-linear response to the electric field, the gradient of the chemical potential and temperature, 
there are also terms that give the response to the higher-order derivatives of the potentials. 
These new, non-local responses, which have not been discussed before, might play an important role for some materials and/or in certain conditions, like extreme miniaturization. 
Our solution provides the general solution of the Boltzmann equation in the relaxation time approximation (or equivalently the particular solution for the specific boundary conditions).  
It differs from the Hilbert expansion which provides only one of infinitely many solutions which may or may not satisfy the required boundary conditions.

\end{abstract}


\maketitle

\section{Introduction}

The phenomenological transport equations for the charge and heat currents are 
at the core of the description of electric and electronic devices of any type.
These equations relate the local charge and heat current densities, $\JJ(\rr)$ and $\JJ_{\En}(\rr)$, 
to the local thermodynamic forces given by the gradients of the electrical potential $\phi$, chemical potential $\mu$, and temperature $T$. 
Often, they are written as 
\begin{align}
	\JJ(\rr)&= \sigma (\EE + \frac{\nabla_{} \mu}{e})-\sigma\alpha \nabla_{} T+ \sigma^{\scriptscriptstyle [\EE^2]} (\EE + \frac{\nabla_{} \mu}{e})^2 \label{eq:phentranspincomplcharge}\\
	\JJ_{\En}(\rr)& = \sigma\alpha T (\EE + \frac{\nabla_{} \mu}{e}) - (\kappa+\sigma\alpha^2T) \nabla_{} T ~, 
	\label{eq:phentranspincomplenergy}
\end{align}
where $\EE=-\nabla \phi$ is the local electric field, and the conductivity $\sigma$, the Seebeck coefficient $\alpha$, 
and the thermal conductivity $\kappa$ are the position-dependent transport coefficients describing the linear part of the response.  
In some cases, the experiments indicate the presence of the non-linear response, but, of all the possible terms, 
we wrote down only the one that is proportional to the square of the electro-chemical force, 
with $\sigma^{\scriptscriptstyle [\EE^2]}$ as the transport coefficient. 

Within the linear response theory, Onsager\cite{1931onsager_1}, Kubo,\cite{kubo_1957} and Luttinger\cite{Luttinger64} 
explained how to relate the coefficients of the driving fields to microscopic quantities
but the microscopic content of $\sigma^{\scriptscriptstyle [\EE^2]} $ is less clear. 
Our aim, within the semi-classical Boltzmann approach, 
is to relate $\sigma^{\scriptscriptstyle [\EE^2]} $ and similar terms in the expansion of the current densities to the energy dispersion and 
the scattering matrix of the electrons.
However, as shown in detail below, the expansion should contain all the powers of the thermodynamic forces and their derivatives, 
i.e., a consistent theory should include all the driving forces up to a given order.  
Beyond the linear order, the current response is usually very small but, under specific conditions, the non-linear driving forces can play a role; 
for instance, when the forces are large or when they exhibit large variations across the sample.

In this article, we show how to obtain a systematic expansion of the current densities in terms of the driving forces.
We find that Eqs.~\eqref{eq:phentranspincomplcharge} and \eqref{eq:phentranspincomplenergy} are incomplete and that non-local terms proportional 
to the higher order derivatives of the potentials, such as $\nabla \EE$, $\partial^2 \mu/\partial \rr^2$ and $\partial^2 T/\partial \rr^2$,  
have to be included as well. 
 This has an obvious fundamental relevance. While some non local effects have been studied so far within different approximations and approaches (as for instance the anomalous skin effect \cite{Reuter.1948,Chambers.1952}), there is no systematic and broad classification of non-local effects, consistent with the definition of other effects like conductivity, Seebeck, Hall, and so on. This consequently leads to neglecting effects which can be important in the description of devices, as for instance the charge current proportional to $\nabla \EE$ in charged bulk regions. 
However, our results are also of a more direct and specific interest for the treatment of modern semiconductor devices. 
The miniaturization requires ever smaller components, with smaller and smaller active regions.\cite{Chau.2007,Iwai.2009,Ferain.2011,Alamo.2011} 
In these regions, the thermodynamic potentials charge enormously over small distances and their higher-order derivatives gain in  importance. 
Thus, even if the prefactors of the new terms are very small, their overall contribution to the total current could be significant 
and they can have an impact on the  performance of the device. 

Similarly, modern thermoelectric devices,\cite{He.2011,Rogl13,Koumoto.13,Heremans.2013,Kanatzidis.2014} 
designed so as to optimize the efficiency under given operating conditions,\cite{mahan.1991,mahan.2000,Seifert,KimLiu} 
are often heterogeneous and have a non-linear distributions of temperature and chemical potential. 
In that case, additional terms are needed to describe the charge and energy current densities.  
An advance along these line requires a quantum mechanical engineering that is hardly possible 
without an insight from  theory\cite{Nolas.2001,ZlaticMonnier,battiatoTomczak} 
 
The theoretical analysis is usually performed in several steps. First, the electronic structure is calculated by density functional theory and the transport coefficients are 
obtained by the linearized Boltzmann equation,\cite{boltzmann} because the Kubo approach is most often too difficult to use for real materials. 
Even the Boltzmann approach requires a number of simplifications, as described in the classic  textbooks.\cite{Ziman_book,Ashcroft1976} 
Once the transport coefficients are known, the currents given by  Eqs.~\eqref{eq:phentranspincomplcharge} and \eqref{eq:phentranspincomplenergy} can be substituted in the continuity equations for the charge  and energy conservation.
Given the appropriate boundary conditions, these equations provide, together with the Poisson equation for the electric field,  
  the  temperature and electrical and chemical potential at every macroscopic point of the sample. 
Thus, one can find the operating conditions, engineer the right composition of the material, and optimize the overall efficiency of the device.

Using the semi-classical Boltzmann theory, and the relaxation time approximation (RTA), 
the present paper formalizes the procedure outlined above and shows how to obtain, in a systematic and rigorous way, 
the transport equations and transport coefficients of inhomogeneous samples. 
Expanding the solution of the Boltzmann equation (BE) in terms of the driving forces, 
we  derive the terms beyond the linear response,  reproduce the terms in Eqs.~\eqref{eq:phentranspincomplcharge} 
and \eqref{eq:phentranspincomplenergy}, 
and show that additional, new terms arise. 
We also show that the effect of the microscopic boundary conditions can be neglected, when the size of the system is large 
with respect to the diffusion length $l=\tau v_{\kk}$.  Here, $\tau$ is the scattering relaxation time, $ v_{\kk}$ is the group velocity of the electron wave packet, 
 and we are referring to the microscopic boundary conditions for the BE, which define the momentum space distribution of incoming electrons at the boundary.

From the mathematical point of view, the proposed expansion has a major advantage over other expansions that are commonly 
used for the BE, like the Hilbert \cite{Hilbert.1917} and Chapman-Enskog expansion\cite{Chapman,Enskog,Cercignani,Chopard}. 
These expansions disregard the boundary conditions i.e., they expand just one of infinitely many possible solutions.
In general, the solution generated by these expansions will not satisfy the boundary conditions 
of the real problem. The expansion proposed in this paper yields, within the RTA, the general solution of the BE. 
This is then used to find the particular solution satisfying the required boundary conditions. 
Thus, we can deal in a consistent way not only with the response to the higher-order driving forces  
but we can also describe the transport properties in the vicinity of the boundary (for instance, the anomalous skin effect \cite{Reuter.1948,Chambers.1952}). 
For simplicity, the present article is restricted to the static case and it addresses in more detail only the former effects.

The paper is organized as follows. Section~\ref{sec:TransportEquations}  introduces the Boltzmann and Poisson equations,  
and the relaxation time approximation, which leads to two further equations required by the energy and particle conservation.  
For a given boundary condition, these four coupled equations determine the electron distribution function, the temperature, the chemical potential, and the electrical field everywhere in the sample. 
Supplementarily, we show in Appendix  \ref{sec:continuityEquations} the equivalence of the equations for the charge and energy conservation 
to the continuity equations for the charge and heat current densities.
In Section~\ref{sec:ExpansionBoltzmann}, rather than solving all these equations simultaneously,  
we focus on the Boltzmann equation treating the temperature, chemical and electrical potential as arbitrary  known functions.  
The central part of the article is the expansion of these arbitrary functions in Taylor series,  which generates an expansion of the non-equilibrium electron distribution function in terms of the driving forces. The coefficients in that expansion satisfy coupled differential equations, as shown  in Section~\ref{sec:driftreactequation}. The problem of setting up the proper boundary conditions and solving these equations is discussed in Section~\ref{sec:boundaryConditions}. 
The major advantage of our expansion over the Hilbert expansion is explained in Appendix~\ref{sec:hilbertexpansion}.  In Section~\ref{sec:macrodevices} we show that, for macroscopic samples, a further approximation can be done 
which leads to major simplifications (this approximation is equivalent to neglecting the effects that are relevant only close to the surface). 
In Section~\ref{sec:macrtransporteq} we derive the generalized transport equations for the heat and charge currents  in the bulk, 
and find additional terms that have not been discussed hitherto.  
Sec.~\ref{sec:examplesemicond} provides one example which shows the influence of the new thermodynamic forces on the behavior of materials: 
we analyze the depletion region in a metal-semiconductor junction. Section~\ref{sec:conclusion} briefly discusses and summarizes our results.

\section{Transport equations} \label{sec:TransportEquations}

Bloch's quantum extension of the Boltzmann's   theory derives the transport properties of a degenerate electron gas
from the distribution function $g(t,\rr,\kk)$, where  $\rr$ and $\kk$  are the coordinates 
of an electron at time $t$ in the real and momentum space, respectively. 
For electrons moving in the presence of a scalar and vector potentials $V(\rr,t)$  and $\AAA(\rr,t)$, 
and confined to a single band, the distribution function satisfies the 
Boltzmann equation \cite{boltzmann,Ziman_book}   (we neglect, for simplicity, the anomalous contribution to the velocity, 
which can easily be included in the BE \cite{Sinitsyn.2008,Xiao.2008})
\begin{align} \label{eq:Boltzmann}
	&\frac{\partial g}{\partial t}+\frac{\nabla_{\kk} \En}{\hbar}     \cdot \nabla_{\rr} g ~ + \\
	 & \frac{e}{\hbar}  \left[ \nabla_{\rr} V + \frac{\partial \AAA}{\partial t}  - \frac{\nabla_{\kk} \En}{\hbar}  \times \left( \nabla_{\rr} \times \AAA \right)\right] \cdot \nabla_{\kk} g =\left(\frac{dg}{dt}\right)_{\!\!col} \nonumber ~.
\end{align}
Here, $\En=\En(\kk)$ is the energy dispersion provided by the band structure calculations, 
$e<0$ is the electron charge, $\hbar$ is the Planck constant,  
and $\left(dg/dt\right)_{\!~\!col}$ is the collision integral which describes 
the change of the distribution function due  to the electron-electron (e-e) and, possibly, 
the electron-phonon (e-ph) scatterings.  
The extension to a multi-band system is straightforward and 
it amounts to a summation over a band index\cite{Ziman_book,Ashcroft1976}.
The charge and current densities defined by $g(t,\rr,\kk)$ have to be compatible 
with the electrodynamic potentials $V(\rr,t)$ and $\AAA(\rr,t)$, as required by the  Maxwell equations.  
The exact solution of the Boltzmann-Maxwell system of equations satisfies all the conservation laws 
compatible with the invariance of the Hamiltonian with respect to the symmetry operations\cite{lifshitz_pitaevski}
  and it determines completely the transport properties of the system. 
Unfortunately, except in the most simple cases, the presence of the collision integral makes the exact solution inaccessible. 
 
In many applications, for example, when engineering an optimal material for a thermoelectric device, 
one tries to infer the transport properties from the available band structure data, neglecting the details 
of the relaxation mechanisms. In that case, the standard approach is to assume  that the scattering drives the system 
towards equilibrium and  to replace the collision integral by a simple expression, 
\begin{equation}\label{eq:relaatimeapprox0}
	\left(\frac{dg}{dt}\right)_{\!\!col} \approx -\frac{g-g_0}{\tau} , 
\end{equation}
where  
$g_0 \left( T, \mu, \En \right)$ is the unperturbed local distribution function 
defined by  local  temperature  $T(\rr)$ and local chemical potential $\mu(\rr)$. 
The assumption that the main effect of the scattering processes is the restoration of local  thermodynamic 
equilibrium on the timescale given by  $\tau$ defines the RTA of the Boltzmann equation.\cite{Ashcroft1976}
Since we are interested in the transport properties of an electron fluid, we choose, 
\begin{equation} \label{eq:relaatimeapprox}
	g_0 (\rr,\kk) =  f_{FD} \left( T (\rr), \mu(\rr), \En(\kk) \right) 
	= 
	\frac{1}{1+e^{\tfrac{\En(\kk)-\mu(\rr)}{k_B T(\rr)}}} ~, 
\end{equation}
where, $k_B$ is the Boltzmann constant and $f_{FD} $  the Fermi-Dirac distribution. 

In what follows, we consider the transport properties in a stationary state, 
such that $\partial g/\partial t=0$, and solve rigorously the static Boltzmann equation using the RTA.  
We take into account the electric field $\EE(\rr)= - \nabla_{} V(\rr)$ but neglect, for simplicity, 
the magnetic field. Thus,  we replace the integro-differential equation Eq.~\eqref{eq:Boltzmann} 
 by a  generalized  drift-reaction (convection-reaction) equation
\begin{equation} \label{eq:Boltzmann_stat}
	\frac{1}{\hbar}   \nabla_{\kk} \En  \cdot \nabla_{\rr} g - \frac{e}{\hbar}  \EE \cdot \nabla_{\kk} g 
	=
	-\frac{g-f_{FD} \left( T\left(\rr\right), \mu\left(\rr\right), \En \left( \kk \right) \right)}{\tau} ~.
\end{equation}
The relaxation time $\tau$ is  treated either as a free parameter, which provides the best fit 
to the experimental data, or it is calculated  in  the perturbation theory \cite{Ziman_book}. 

Unlike the exact solution of Eq.~\eqref{eq:Boltzmann}, the solution of Eq.~\eqref{eq:Boltzmann_stat} 
does not automatically satisfy the fundamental conservation laws, like the particle number and the energy conservation.  
To make the RTA  physically acceptable we enforce the local particle and energy conservation 
by constraining the functions $T=T(\rr)$ and $\mu=\mu(\rr)$.
The conservation of the local particle density $n\left(\rr \right) =\int g \left(\rr,\kk \right) d^3k$ 
follows from the requirement 
\begin{equation} \label{eq:number_cons}
\int g \left(\rr,\kk \right) d^3k = \int f_{FD} \left( T\left(\rr\right), \mu\left(\rr\right), \En \left(\kk \right) \right) d^3k 
\end{equation}
while the conservation of the total energy density of interacting electrons is enforced by the equation  
\begin{equation}
\begin{split} \label{eq:energy_cons_0}
	\int \!\En\left( \kk \right)\, g \left(\rr,\kk \right) d^3k =\!\!
	 \int\! \En\left( \kk \right)\, f_{FD} \left( T\left(\rr\right)\!, \mu\left(\rr\right)\!, \En \left( \kk \right) \right) d^3k.
\end{split}
\end{equation}
If the electrons scatter on some additional degrees of freedom, like phonons, 
the scattering process changes their energy by  $ \Delta \epsilon_{e-ph} $ 
which has to be added to the right-hand-side of Eq.~\eqref{eq:energy_cons_0}. 
The consistency of the charge density and the electrical field is enforced by the Poisson equation, 
\begin{equation}
\nabla_{} \EE(\rr,t) = \frac{e\, n(\rr) + \rho_{\mbox{ion}}(\rr)}{\epsilon_0} ~, 
\label{Eq:Poisson}
 \end{equation}
where  $\rho_{\mbox{ion}}(\rr)$ is the background charge that ensures the overall charge neutrality. 
The self-consistent solution of Eqs.~\eqref{eq:Boltzmann_stat} - \eqref{Eq:Poisson} 
provides $ g \left(\rr,\kk \right) $, $\EE(\rr)$, $T(\rr)$ and $\mu(\rr)$ at every point in the sample. 

As shown in Appendix \ref{sec:continuityEquations}, the conservation of charge and energy imply the continuity equations for the charge and energy current densities. 
In a stationary state (and in the absence of electron-phonon scatterings), the  current densities satisfy 
\begin{align} \label{eq:div_charge_current}
	&  \nabla_{}\cdot \JJ(\rr)=0  \\
	& \nabla_{}\cdot \JJ_{\En}(\rr)=W~,
	\label{eq:div_enenrgy_current}
\end{align}  
where the charge and energy current densities are 
\begin{eqnarray} \label{eq:charge_current}
	\JJ \left(\rr\right) &=&   \frac{e}{\hbar}   \int      \; \nabla_{\kk} \En \;  g\left(\rr,\kk \right)  d^3k ~, \\
	\JJ_\En \left(\rr\right) & =& \ \frac{e}{\hbar}  \int {\nabla_{\kk} \En} \; \En  \;  g\left(\rr,\kk \right) d^3k ~,
	 \label{eq:energy_current}
\end{eqnarray}
and $W$ is the work done by the applied electric field per unit time
\begin{eqnarray} \label{eq:work}
	W\left(\rr\right)&=&  \frac{e}{\hbar}  \int \nabla_{\kk} \En \cdot  \EE\left(\rr \right) \;   g\left(\rr,\kk \right) \,d^3k= \JJ \cdot \EE.	
\end{eqnarray}

Thus, finding the electron distribution function in the RTA implies solving  the Boltzmann  
equation~\eqref{eq:Boltzmann_stat} for $g(\rr,\kk)$, together with the continuity equations \eqref{eq:div_charge_current} 
and \eqref{eq:div_enenrgy_current}, and the Poisson equation \eqref{Eq:Poisson}. Equivalently one can solve the Boltzmann  
equation~\eqref{eq:Boltzmann_stat} for $g(\rr,\kk)$ and Eqs.~\eqref{Eq:Poisson}  -- \eqref{eq:energy_current} for five unknown quantities $\JJ \left(\rr\right), \JJ_\En \left(\rr\right), T \left(\rr\right), \mu\left(\rr\right)$   and $\EE(\rr)$.

In general, the above equations have  infinitely many solutions and the physically relevant one is defined by the specific boundary conditions which provide  
$g (\rr_B,\kk)=g_B(\rr_B,\kk)$ at every point $\rr_B$ of the boundary.  
Note the difference between these detailed, microscopic boundary conditions and the one which specifies just the macroscopic quantities, 
like temperature and electrical and chemical potentials, at the interfaces. 
The microscopic boundary conditions for the Boltzmann equation specify the momentum distribution of the electrons coming from the neighboring material, 
and take also into account the reflection and scattering of the incoming electrons at the interface. 
In addition to the temperature of the injection, they should provide, for example, a detailed information on the band structure of the neighboring material, 
the $\kk$-dependent injection probability or reflectivity. The construction of boundary conditions that determines a specific physical situation at the interface 
is a non-trivial problem which is not addressed further in this work. In the following, we simply assume  the boundary conditions to be known.  

\section{Expansion of the Boltzmann  distribution function in terms of the generalized forces} 
\label{sec:ExpansionBoltzmann}

Although the Boltzmann equation simplifies considerably within the RTA, 
solving Eq.~\eqref{eq:Boltzmann_stat} for $g(\rr,\kk)$, together with Eqs.~\eqref{eq:number_cons} -- \eqref{Eq:Poisson}  
for $T(\rr)$, $\mu(\rr)$, and $\EE(\rr)$, is still a formidable task. 
To solve it, we integrate Eq.~\eqref{eq:Boltzmann_stat} for completely arbitrary functions $\mu(\cdot)$, $T(\cdot)$, and $E(\cdot)$, 
using an expansion of $g(\rr,\kk)$ in terms of the forces that arise out of equilibrium and 
to which the system responds by setting up the currents.
Such approximations, based on physical arguments and irreversible thermodynamics, are often used but, here, we 
present a systematic expansion which clarifies the range of validity of the textbook solutions and identifies the new driving forces. 

For a given microscopic boundary condition ($BC$), the solution of Eq.~\eqref{eq:Boltzmann_stat} has a unique value at every point $\{\rr,\kk\}$ of the phase space, so that  $g(\rr,\kk)$  is a functional $g_F$
defined on functions $\mu(\cdot)$, $T(\cdot)$ and $\EE(\cdot)$, and the BC themselves. That is, 
\begin{equation} \label{eq:operator1}
	g(\rr,\kk)=g_F \left[\rr, \kk, T(\cdot) ,\mu(\cdot),\EE(\cdot),BC\right] ~.
\end{equation}
Assuming $\mu(\cdot)$, $T(\cdot)$ and $\EE(\cdot)$ are analytic functions, we expand them into Taylor series around point $\rr$ 
and treat $g(\rr,\kk)$, without any loss of information\cite{functional}, not as a functional but as a function of infinitely many variables,  
\begin{align} \label{eq:operator2}
	g(\rr,\kk)=\tilde{g} \big(&\rr,\kk,  T(\rr),\mu(\rr), \EE(\rr) , \nabla T(\rr),\nabla \mu(\rr),\\
	& \nabla \EE(\rr), \frac{\partial^2 T(\rr)}{\partial \rr^2},\frac{\partial^2 \mu(\rr)}{\partial \rr^2},\frac{\partial^2 \EE(\rr)}{\partial \rr^2}, ..., BC\big) .\nonumber
\end{align}
 (Operator $\nabla$ denotes $\nabla_{\rr}$, whenever the function operated on depends solely on the position $\rr$ and no ambiguity can arise.)
Obviously, analytic functions $T(\cdot) ,\mu(\cdot),\EE(\cdot)$ are completely defined by their values and the values of all their derivatives 
at any point of the sample. For example, $\mu(\cdot)$ is determined everywhere in its region of definition, if we provide, at point $ \rr$, 
 the values $\mu(\rr)$, $\nabla_{} \mu(\rr)$, $\partial^2 \mu(\rr)/\partial \rr^2$, 
$\partial^3 \mu(\rr)/\partial \rr^3$, etc. 
If one is interested in the charge transport close to the interface, say, to model the Kapitza resistance \cite{Pollack.1969}, 
the interface defines a discontinuity and the Taylor extension cannot be used. 
To circumvent that problem, one can split the system into two halves, one to the left and one to the right of the boundary, 
and use separate Taylor expansions on each sides of the interface. 
However, solving for such boundary conditions, including the Kapitza resistance, becomes cumbersome.

To proceed, we introduce the vector 
$\vec{\xi}=\{\xi_1,\xi_2,\xi_3,...\}$, where $\xi_1=\rr$, $\xi_2=\kk$, $\xi_3=T$, $\xi_4=\mu$, 
$\xi_5= \EE$, $\xi_6=\nabla_{} T$, etc. The components ${\xi_i}$, for $i\geq 5$,  describe 
the driving forces to which the system responds.
This notation indicates that even though $\tilde{g}(\vec{\xi}\,)$ is  defined on an infinite dimensional vector space,  
only the values assumed by $\tilde{g}(\vec{\xi}\,)$ on a small subspace of the whole definition space 
are physically relevant. In particular, we are interested in $\tilde{g}(\vec{\xi}\,)$ on the hypersurface 
defined by $\xi_1=\rr$, $\xi_2=\kk$, $\xi_3=T(\rr)$, $\xi_4=\mu(\rr)$, $\xi_5=\EE(\rr)$, 
$\xi_6=\nabla_{} T(\rr)$, $\xi_7=\nabla_{} \mu(\rr)$ etc.

Next, we expand  $\tilde{g}(\vec{\xi}\,)$ in a Taylor series 
 around the point $\vec{\xi}_0=\{\xi_1,\xi_2,\xi_3,\xi_4,0,0,0,\ldots\}$, i.e., 
we expand $\tilde{g}(\vec{\xi}\,)$ with respect to the variables $\xi_i$ around $\xi_i=0$, for all $i\geq 5$. 
Thus, we write 
\begin{align} \label{eq:expansion}
	&\tilde{g}(\vec{\xi}\,)= g^{[0]} [\xi_1,\xi_2, \xi_3, \xi_4]   \nonumber\\
	&+\;\;\delta g^{[ \xi_5]} [\xi_1,\xi_2, \xi_3, \xi_4]\; \xi_5  \nonumber\\
	&+\;\;  \delta g^{[\xi_6 ]} [\xi_1,\xi_2, \xi_3, \xi_4]\; \xi_6    \nonumber\\
	&+\;\;  \delta g^{[\xi_7 ]} [\xi_1,\xi_2, \xi_3, \xi_4]\; \xi_7    \\
	&+\;\;  \delta g^{[ \xi_5^2]} [\xi_1,\xi_2, \xi_3, \xi_4]\; \xi_5^2  \nonumber \\
	&+\;\;  \delta g^{[\xi_8]} [\xi_1,\xi_2, \xi_3, \xi_4]\; \xi_8   \nonumber\\
	&+\;\;  \delta g^{[\xi_8^2 ]} [\xi_1,\xi_2, \xi_3, \xi_4]\; \xi_8^2   \nonumber\\
	&+\;\;  \delta g^{[\xi_5 \xi_{6}]} [\xi_1,\xi_2, \xi_3, \xi_4] \;\xi_5 \xi_{6}+ ...  \nonumber 
\end{align}
where the coefficients $ \delta g^{[\alpha,\beta,\gamma,....]} [\xi_1,\xi_2, \xi_3, \xi_4] $ 
depend on the first four variables and the boundary conditions. 
{(The explicit dependence on the boundary conditions has been omitted, for brevity.)}
In terms of the physically more transparent symbols, we have
\begin{align} \label{eq:expansion_subst}
	g(\rr,\kk) &= g^{[0]} [\rr,\kk, T(\rr), \mu(\rr)]  \nonumber\\
	&\;\;\;\; +\delta g^{[ \EE]} [\rr,\kk, T(\rr), \mu(\rr)]  \,\EE(\rr)+ ...  \\
	&\;\;\;\; +\delta g^{[ \nabla_{} T]} [\rr,\kk,  T(\rr), \mu(\rr)] \,\nabla_{} T(\rr)  \nonumber\\
	&\;\;\;\;+ \delta g^{[ \nabla_{} \mu]} [\rr,\kk,  T(\rr), \mu(\rr)] \,\nabla_{} \mu(\rr) \nonumber\\
	&\;\;\;\;+ \delta g^{[ \EE^2]} [\rr, \kk, T(\rr), \mu(\rr)]\, (\EE(\rr))^2+ ... \nonumber\\
	&\;\;\;\;+ \delta g^{[ \partial^2 T/\partial \rr^2 ]} [\rr, \kk, T(\rr), \mu(\rr)] \, \partial^2 T(\rr)/\partial \rr^2+ ... \nonumber\\
	&\;\;\;\;+\delta g^{[ (\partial^2 T/\partial \rr^2)^2 ]} [\rr, \kk, T(\rr), \mu(\rr)] \,(\partial^2 T(\rr)/\partial \rr^2)^2 \nonumber\\
	&\;\;\;\;+ \delta g^{[ \EE, \nabla_{} T ]} [\rr, \kk, T(\rr), \mu(\rr)] \, \EE(\rr) \nabla_{} T(\rr) + ... ,\nonumber
\end{align}
where the (still unknown) coefficients  $\delta g^{[E]}$, $\delta g^{[ \nabla_{} T]}$, $\delta g^{[ \nabla_{} \mu]}$, etc.~
describe the change in the distribution function due to the applied forces $\EE$, $\nabla_{} T$, 
$\nabla_{} \mu$, etc.  These coefficients can be computed by substituting $\tilde{g}(\vec{\xi}\,)$, given by Eq.~\eqref{eq:expansion_subst}, 
into  the Boltzmann equation ~\eqref{eq:Boltzmann_stat} and collecting the terms to order $\EE(\rr)$, 
$\nabla_{} T(\rr)$, $\nabla_{} \mu(\rr)$, $ \partial^2 T(\rr)/\partial \rr^2$, 
$\EE(\rr)^2$, $\EE(\rr) \nabla_{} T(\rr)$, $\EE(\rr) \nabla_{} \EE(\rr)$,  etc. 
The first term in Eq.~\eqref{eq:Boltzmann_stat} yields (we only show the first few terms):
\begin{widetext}
\begin{align}   \label{eq:expansion_subst2}
	&\nabla_{\rr} g(\rr,\kk) 
	= \nabla_{\rr} \left[  g^{[0]} [\rr,\kk, T(\rr), \mu(\rr)]   
	 +  \delta g^{[ \EE]} [\rr,\kk, T(\rr), \mu(\rr)] \, \EE(\rr)  \right.\nonumber \\
	 &\;\;\;\;\;\;\;\;\;\;\;\;\;\;\;\;\;\;\;\;\;\;\;\;\;\;\;\;+  \delta g^{[ \nabla_{} T]} [\rr,\kk,  T(\rr), \mu(\rr)] \,\nabla_{} T(\rr) 
	 +  \delta g^{[ \nabla_{} \mu]} [\rr,\kk,  T(\rr), \mu(\rr)] \,\nabla_{} \mu(\rr)  \nonumber\\
	&\;\;\;\;\;\;\;\;\;\;\;\;\;\;\;\;\;\;\;\;\;\;\;\;\;\;\;\; + \left.\delta g^{[ \EE^2]} [\rr, \kk, T(\rr), \mu(\rr)] \,(\EE(\rr))^2 
	+  \delta g^{[ \EE,\nabla_{} \mu]} [\rr, \kk, T(\rr), \mu(\rr)] \,(\EE(\rr))\nabla_{} \mu+ ...  \right]  \nonumber\\
       &= 
       \nabla_{\rr} g^{[0]} [\rr,\kk, T, \mu ]  + \frac{\partial g^{[0]} [\rr,\kk, T, \mu ]}{\partial T}  \nabla_{} T + \frac{\partial g^{[0]} [\rr,\kk, T, \mu ]}{\partial \mu}\nabla_{} \mu  \\
	&\;\;\;+\nabla_{\rr} \delta g^{[ \EE]} [\rr,\kk, T, \mu ] \,\EE  + \frac{\partial \delta g^{[ \EE]} [\rr,\kk, T, \mu ] }{\partial T} \EE \, \nabla_{} T + \frac{\partial \delta g^{[ \EE]} [\rr,\kk, T, \mu ] }{\partial\mu} \EE\, \nabla_{} \mu + \delta g^{[ \EE]} [\rr,\kk, T, \mu ] \, \nabla_{} \EE  \nonumber \\
	&\;\;\; +\nabla_{\rr} \delta g^{[ \nabla_{} T]} [\rr,\kk, T, \mu ] \,\nabla_{} T  + \frac{\partial \delta g^{[ \nabla_{} T]} [\rr,\kk, T, \mu ] }{\partial T} (\nabla_{} T)^2 + \frac{\partial \delta g^{[ \nabla_{} T]} [\rr,\kk, T, \mu ] }{\partial \mu}  \nabla_{} T\, \nabla_{} \mu + \delta g^{[\nabla_{} T ]} [\rr,\kk, T, \mu ] \, \frac{\partial^2 T}{\partial \rr^2}  \nonumber \\
       &\;\;\; +\nabla_{\rr} \delta g^{[ \nabla_{} \mu]} [\rr,\kk, T, \mu ] \,\nabla_{} \mu  + \frac{\partial \delta g^{[ \nabla_{} \mu]} [\rr,\kk, T, \mu ] }{\partial \mu} (\nabla_{} \mu)^2 + \frac{\partial \delta g^{[ \nabla_{} \mu]} [\rr,\kk, T, \mu ] }{\partial T}  \nabla_{} T\, \nabla_{} \mu + \delta g^{[\nabla_{} \mu ]} [\rr,\kk, T, \mu ] \, \frac{\partial^2 \mu}{\partial \rr^2}  \nonumber \\
	&\;\;\; +\nabla_{\rr} \delta g^{[ \EE^2]} [\rr,\kk, T, \mu ] \,\EE^2  + \frac{\partial \delta g^{[ \EE^2]} [\rr,\kk, T, \mu ] }{\partial T} \nabla_{} T\, \EE^2 + \frac{\partial \delta g^{[ \EE^2]} [\rr,\kk, T, \mu ] }{\partial \mu} \EE^2\, \nabla_{} \mu +2 \delta g^{[\EE^2 ]} [\rr,\kk, T, \mu ] \, \EE \,\nabla_{} \EE  \nonumber\\
	&\;\;\;+ .... \nonumber 
\end{align}
\end{widetext}
where the effect of $\nabla_{\rr}$, operating on composite functions in the first equation, 
is computed using the usual rules for the derivatives of the composite function, 
while $\nabla_{\rr} g^{[ 0]} [\rr,\kk, T, \mu ]$, etc.~, in the second equation, denotes the derivative of $g^{[0]} [\rr,\kk, T, \mu]$ with respect to the first variable only. 

The terms obtained by substituting Eq.~\eqref{eq:expansion_subst}  in the second  
and the third term of the Boltzmann Eq.~\eqref{eq:Boltzmann_stat} are easily written down 
and are not reported separately.  The sum of all three terms can be written as
\begin{widetext}
\begin{equation}  \label{eq:Boltzmann_stat_expanded}
\begin{split}
	\frac{1}{\hbar}   \nabla_{\kk} \En  \cdot \nabla_{\rr} g - \frac{e}{\hbar}  \EE \cdot \nabla_{\kk} g 
	&+\frac{g-f_{FD} \left( T\left(\rr\right), \mu\left(\rr\right), \En \left( \kk \right) \right)}{\tau} =  \\
	=~~~~~~&\left(\frac{1}{\hbar}   \nabla_{\kk} \En  \cdot \nabla_{\rr} g^{[0]} 
	                   +\frac{g^{[0]}-f_{FD} \left( T, \mu, \En\right)}{\tau} \right) \\
	&+ \left( \frac{1}{\hbar}   \nabla_{\kk} \En  \cdot \nabla_{\rr} \delta g^{[\nabla_{} T]} +\frac{1}{\hbar}  \frac{\partial g^{[0]} }{\partial T}  \nabla_{\kk} \En  +\frac{\delta g^{[\nabla_{} T]}}{\tau} \right) \nabla_{} T  \\
	&+ \left( \frac{1}{\hbar}   \nabla_{\kk} \En  \cdot \nabla_{\rr} \delta g^{[\nabla_{} \mu]} 
	  +\frac{1}{\hbar}  \frac{\partial g^{[0]} }{\partial \mu}  \nabla_{\kk} \En  +\frac{\delta g^{[\nabla_{} \mu]}}{\tau} \right) \nabla_{} \mu  \\
	& +\left( \frac{1}{\hbar}   \nabla_{\kk} \En  \cdot \nabla_{\rr} \delta g^{[\EE]}  
	- \frac{e}{\hbar}  \mathbf{u}_{\EE}  \cdot \nabla_{\kk} g^{[0]} +\frac{\delta g^{[\EE]}}{\tau} \right) \EE + ... \\ =0 
	 ~, 
\end{split}
\end{equation}
\end{widetext}
where  $\mathbf{u}_{\EE}$ is the unit vector in the direction of the electric field 
and only the linear terms are shown, because the structure of the higher order terms is obvious. 
Since $T(\rr)$, $\mu(\rr)$ and $\EE(\rr)$ are arbitrary functions, 
the above equation can only be satisfied if all the brackets vanish. We therefore have to impose that all the expressions within brackets in Eq.~\ref{eq:Boltzmann_stat_expanded} have to vanish separately.

Thus, we reduced the Boltzmann equation to an infinite sequence of coupled differential equations 
which describe the change in the distribution function in response to the driving forces. 
Each equation specifies a particular response function 
 $\delta g^{[\alpha]}[\rr,\kk, T, \mu ] $ (where $T$ and $\mu$ are treated as variables)
which corresponds to a particular driving force $\alpha$ 
and  the differential operator in these equation is operating on the first variable of $\delta g^{\alpha}[\xi_1, \xi_2, \xi_3,\xi_4]$ only. 
The solution can be constructed sequentially, starting from the lowest order   
and specifying, for every equation, a particular boundary condition regarding the variable $\rr$. 
The  construction has to ensure that the sum of all the contributions yields  $g(\rr,\kk)$ 
which satisfies the boundary condition imposed on the solution of Eq.~\eqref{eq:Boltzmann_stat}  (more on this in Sec.~\ref{sec:boundaryConditions}).

\subsection{Drift-reaction equations } \label{sec:driftreactequation}

We now discuss  the zeroth and the first order distribution function defined 
by the expansion Eq.~\eqref{eq:Boltzmann_stat_expanded} 
and relate them to what is known from the literature. 
We also provide a few typical examples of the higher order terms. 
The zeroth-order distribution function is obtained by setting to zero the first bracket 
in Eq.~\eqref{eq:Boltzmann_stat_expanded},  which gives 
\begin{equation} \label{eq:zero_order}
	\frac{1}{\hbar}   \nabla_{\kk} \En  \cdot \nabla_{\rr} g^{[0]} +\frac{g^{[0]}-f_{FD} \left( T, \mu, \En\right)}{\tau} =0~,
\end{equation}
where $\nabla_{\rr} g^{[0]} $  is again the derivative  of $g^{[0]} [\rr,\kk, T, \mu]$ with respect to its first variable.
The spatial part of Eq.~\eqref{eq:zero_order} is a convection-reaction equation and  the solution requires   
the value of $g^{[0]}[\rr,\kk,T,\mu]$ on the boundary. 

The change in the distribution function due to an applied electric field $\EE$, is defined by the equation 
(see Eq.~\eqref{eq:Boltzmann_stat_expanded})
\begin{equation} \label{eq:drift_react_2}
	\frac{ \nabla_{\kk} \En}{\hbar}    \cdot \nabla_{\rr} \delta g^{[\EE]} 
	- \frac{e}{\hbar}  \frac{\EE}{\left|\EE\right|} \cdot  \nabla_{\kk} g^{[0]} + \frac{\delta g^{[\EE]}}{\tau}   
	=0~.
\end{equation}
Similarly, the response to a thermal force is obtained by collecting all the first order terms in $\nabla_{} T$,  
which gives 
\begin{equation} \label{eq:drift_react_2T}
	\frac{ \nabla_{\kk} \En}{\hbar}  \cdot  \nabla_{\rr} \delta g^{[\nabla_{} T]} + 
	\frac{ \nabla_{\kk} \En}{\hbar} \frac{\partial g^{[0]}}{\partial T} + \frac{\delta g^{[\nabla_{} T]}}{\tau}   =0 ~;
\end{equation}
the response to a diffusion force  $\nabla_{} \mu$ is given by
\begin{equation} \label{eq:drift_react_2mu}
	 \frac{ \nabla_{\kk} \En}{\hbar}    \cdot \nabla_{\rr} \delta g^{[\nabla_{} \mu]} + \frac{ \nabla_{\kk} \En}{\hbar} \frac{\partial g^{[0]}}{\partial \mu}+ \frac{\delta g^{[\nabla_{} \mu]}}{\tau}   =0.
\end{equation}

Equations \eqref{eq:zero_order} -- \eqref{eq:drift_react_2mu},   without the first  term,  
yield the linear corrections to the Boltzmann distribution function, which is the same as in most textbooks \cite{Ziman_book,Ashcroft1976}.  
These approximate expressions agree also with the results obtained, for instance, by the Hilbert expansion of Eq.~\eqref{eq:Boltzmann_stat} (see Appendix \ref{sec:hilbertexpansion}).
At this stage, it is not obvious that $\nabla_{\rr} \delta g^{[\alpha]}$ can be neglected but, 
in Section \ref{sec:macrodevices}, we show that the rigorous solution of equations \eqref{eq:zero_order} --  \eqref{eq:drift_react_2mu} indeed
assumes the textbook form sufficiently far away from the boundaries. 

The higher order response follows straightforwardly from the expansion Eq.~\eqref{eq:Boltzmann_stat_expanded} 
and yields the terms of two basic types. The first type describes the non-linear response due to the higher 
powers of the gradients of potentials (like $ (\nabla_{} \mu)^2$, $\EE^2$, $(\nabla_{} T)^2$, etc.). 
Many such terms have previously been discussed in the literature\cite{Ziman_book,Ashcroft1976}. 
The second order response to the diffusion force, $ (\nabla_{} \mu)^2$, is defined by the equation 
\begin{equation}\label{eq:drift_react_dmu2}
		\frac{ \nabla_{\kk} \En}{\hbar}    \cdot \nabla_{\rr} \delta g^{[(\nabla_{} \mu)^2]} + \frac{ \nabla_{\kk} \En}{\hbar}  \frac{\partial \delta g^{[\nabla_{} \mu]}}{\partial \mu} + \frac{\delta g^{[(\nabla_{} \mu)^2]}}{\tau}   =0 ~
\end{equation}
and the response to $(\nabla_{} T)^2$ is similar. 

The terms of the second type describe the response to the higher order derivatives of the potentials (like 
$ \partial^2 \mu(\rr)/\partial \rr^2$ or $\nabla_{} E$, etc.) and all their powers. 
These terms have not been considered before, even though they can be comparable to the non-linear terms 
of the same (and higher) order. 
For example, the response to the second derivative of the chemical potential 
$ \partial^2 \mu(\rr)/\partial \rr^2$ reads
\begin{equation} \label{eq:drift_react_d2mu}
	\frac{ \nabla_{\kk} \En}{\hbar}    \cdot \nabla_{\rr} \delta g^{[\frac{\partial^2 \mu}{\partial \rr^2}]} + \frac{ \nabla_{\kk} \En}{\hbar}  \delta g^{[\nabla_{} \mu]} + \frac{\delta g^{[\frac{\partial^2 \mu}{\partial \rr^2}]}}{\tau}   =0.
\end{equation}

A multidimensional Taylor expansion generates also a large number of mixed terms.  
The equations for other higher order responses (say, the one proportional to $ (\nabla_{} \mu)^3$ or $E^3$) or the equations 
for the coefficients depending on the higher derivatives of the fields (like $ \partial^3 \mu(\rr)/\partial \rr^3$), 
or all the cross terms, are obtained by straightforward but tedious calculations.  

The distribution function is now obtained by solving successively equations ~\eqref{eq:zero_order}, \eqref{eq:drift_react_2}, \eqref{eq:drift_react_2T}, etc. 
The structure of these equations is always the same and we can write in the $n$-th order
\begin{align} \label{eq:drift_nth_order} 
\nabla_{\xi_1} \delta g^{[\xi_n]}[\xi_1,\xi_2,\xi_3,\xi_4] & +F_n(\delta g^{[\xi_0]},\delta g^{[\xi_6]},\ldots,\delta g^{[\xi_{n-1}]})  \nonumber \\
	& +\frac{ \delta g^{[\xi_n]}}{\tau} =0,
\end{align}
where $F_n(\delta g^{[\xi_0]},\delta g^{[\xi_6]},\ldots,\delta g^{[\xi_{n-1}]})$ is a known function obtained from the solution of the lower-order equations.

\subsection{Expansion of the boundary conditions} \label{sec:boundaryConditions}

The boundary conditions for Eq.~\eqref{eq:Boltzmann_stat}, written as $g(\rr_B,\kk)=g_B(\rr_B,\kk)$, define how the electrons are injected in the region under  consideration.  For instance a surface subject to an electron flux from vacuum (as in the case of inverse photoemission) is subject to an electron current with an electronic distribution at the entrance that depends on the energy distribution of the injected electrons. Another interesting case is that of a current flowing from a semiconductor into a metal. The injected electrons do not have the same energy distribution as in the case when they are excited by an electric field inside the metal.
In general, from a mathematical point of view.  the population of electrons in the k-space can be described at the boundary $\rr=\rr_B$
by any function of momentum, not necessarily by the Fermi-Dirac distribution or the non-equilibrium distribution which is giving rise to stationary currents. 

A unique solution of  Eqs.~\eqref{eq:zero_order}, \eqref{eq:drift_react_2}, \eqref{eq:drift_react_2T}, etc.~requires a boundary condition in every order. 
Since the differential operator in the drift-reaction equation  for  $\delta g^{[\xi_n]}$ is operating on the the first argument  
of  $\delta g^{[\xi_n]}[\rr,\kk, T, \mu]$,  the most general form of the boundary conditions for the drift-reaction equations is
\begin{equation}\label{eq:boundary_condalpha}
	\delta g^{[\xi_n]}[\rr_{B},\kk, T, \mu]  =\delta g_B^{[\xi_n]}[\rr_{B},\kk, T, \mu],
\end{equation}
where  $\delta g_B^{[\xi_n]}$ specifies the value of $\delta g^{[\xi_n]}$ at the boundary for any given $\kk$,  $T$, and $\mu$. 

Note,  any choice of boundary conditions for the drift-reaction equations is acceptable,  
as long as the sum of all terms in Eq.~\eqref{eq:expansion_subst} yields the correct boundary condition for 
the solution of the Boltzmann equation~\eqref{eq:Boltzmann_stat}. That is, the boundary conditions for the drift-reaction equations 
have to satisfy the supplementary condition 
\begin{align} \label{eq:cond_boundary_cond}
	g_B(\rr_B,\kk) &=g_B^{[0]}[\rr_B,\kk,T(\rr_B),\mu(\rr_B)] \\
	& \!\!\!\!\!\!\! + \sum \delta g_B^{[\xi_n]}[\rr_B,\kk,T(\rr_B),\mu(\rr_B)] \;\xi_n(\rr_B). \nonumber
\end{align}
As long as Eq.~\eqref{eq:cond_boundary_cond} holds,  the sum of all terms in Eq.~\eqref{eq:expansion_subst} 
gives the particular solution of Boltzmann equation ~\eqref{eq:Boltzmann_stat}  which satisfies the required boundary condition 
(assuming the series expansion converges). 

This concludes the construction of the expansion of the solution of the BE in the RTA. 
We emphasize that our expansion of the distribution function takes into account the boundary conditions. 
This implies that our solution yields not just the response to the higher driving forces or the non-local effects in the bulk, 
but it can also describe the effects caused by the specific choice of the boundary conditions  (for instance, our solution can be used 
to discuss the anomalous skin effect). 
In Appendix \ref{sec:hilbertexpansion} we show why this cannot be achieved by other, often used, expansions, as for instance the Hilbert expansion.

\section{Solution of the drift-reaction equation in a macroscopic sample} \label{sec:macrodevices}

In the rest of this paper we focus on the high order responses and non-local effects in the bulk, 
leaving the description of surface effects to future work (by surface effects, we mean the features of the solution that depend on the specific form of the boundary conditions). 
 We will proceed in two  steps. 
First we show that due to the dissipative nature of the scattering term in the RTA,  a specific shape of the BC modifies the solution only close to the boundary. 
This implies that, for a given functional form of $T(\rr)$, $\mu(\rr)$ and $\EE(\rr)$, any BC imposed on the BE leads to the same solution in the bulk. 

For a very small device, 
the full set of the drift-reaction equations~\eqref{eq:zero_order}, \eqref{eq:drift_react_2}, \eqref{eq:drift_react_2T}, etc., has to be solved 
for a given choice of the boundary conditions, subject to the supplementary condition Eq.\eqref{eq:cond_boundary_cond}. 
This rises the problem of the optimal distribution of the boundary conditions among various drift-reaction equations.  
Similar issues arise also in  larger systems close to the physical boundary. However, these problems are not addressed here. 

To understand the extent in which the details of the BC affect the solution far away from the boundary,  
we study the lowest-order drift-reaction equation, Eq.~\eqref{eq:zero_order}, in one dimension. The solution can be written as 
\begin{equation} \label{eq:1dim_convection}
\begin{split}
	&g^{[0]} [x,\kk, T, \mu]=  \\
	&\;\;\;\;\;f_{FD} \left( T, \mu, \En \left( \kk \right) \right) + 
	C(\kk,T,\mu) e^{-\tfrac{\hbar\, x}{\tau  \partial_{k_x} \En}}
\end{split}
\end{equation}
where $\kk, T, \mu$ are arbitrary and the coefficient $C(\kk,T,\mu)$  is defined by the BC at $x=0$. 
Obviously,  the effect of the boundary on the solution $g^{[0]} [x,\kk, T, \mu]$ decreases exponentially 
with the distance from the boundary.  The characteristic decay length is given by 
the mean free path, $l=\tau  v_k$.  
For $x\gg l$, the effect of the boundary is obliterated by the scattering, so that the particular value of $g^{[0]} [x,\kk, T, \mu]$ 
at $x=0$ becomes irrelevant at distances which are much larger than the mean free path.  This feature also holds 
in higher dimensions and for every term in the expansion. 

Therefore, if we are interested only in the bulk solution, which holds far from the boundary, we are free to choose the BC as we like; 
the difference between the true particular solution and the one obtained for a different BC vanishes far away from the boundary. 
This implies that the solution of the Boltzmann equation in the bulk of the sample, is completely determined by the local  temperature, 
chemical potential and electric field, regardless of the microscopic BC.  

As regards the response of the entire sample, the larger the system, the less important the region close to the boundary. 
In a macroscopic device, an accurate treatment of microscopic boundary conditions gives only a very small correction to the response functions  
but it  increases dramatically the computational complexity. 
Hence, we choose the boundary condition so as to minimize the computational efforts; for large enough systems, the error of using such a solution is insignificant.  
Note the difference between the  microscopic boundary conditions for the  BE and the macroscopic boundary conditions for the continuity equations providing
the thermodynamic variables $T$ and $\mu$. The conservations laws given by Eqs.~\eqref{eq:number_cons} and \eqref{eq:energy_cons_0} require 
$T$ and $\mu$ to assume the boundary values specified by the reservoirs. 
The microscopic boundary conditions, imposed on the Boltzmann equation, provide an information on the state of the electrons at the boundary. 

If we decide to disregard the surface effects, we can choose  the BC for   Eq.~\eqref{eq:zero_order}  
as $g^{[0]}[0,\kk,T,\mu]\!=\!f_{FD} \left( T, \mu, \En \left( \kk \right) \right)$, which yields  $C(\kk,T,\mu)=0$ and  
makes the function $g^{[0]}$ independent of the variable $x$. 
The differential operator in Eq.~\eqref{eq:zero_order} can now be dropped and the solution of the  zeroth order drift-reaction 
equation becomes 
\begin{equation} \label{eq:approx0th}
	g^{[0]}[\rr,\kk,T,\mu]=f_{FD} \left( T, \mu, \En \left( \kk \right) \right) ~
\end{equation}
everywhere in the sample. This solution deviates from the exact one only very close to the boundary, where the exponential term cannot be neglected. 

Similarly, if we chose the BC for Eq.~\eqref{eq:drift_nth_order} as 
\begin{equation}
	\delta g^{[\xi_n]}[\rr_B,\kk,T,\mu]= 
	- \tau F_n(\delta g^{[\xi_0]},\delta g^{[\xi_6]},\ldots,\delta g^{[\xi_{n-1}]})~, 
\end{equation}
the $n$-th order solution can be computed analytically and it will be independent on $\xi_1$. 
The set of approximate equations obtained in such a way coincides with the equations generated  
by the Hilbert expansion in the static approximation (see Appendix \ref{sec:hilbertexpansion}).

The approximate solution of the BE  which works very well for bulk materials 
is obtained by summing up all the solutions of the drift-reaction equations. The ensuing   distribution 
function reads
\begin{equation} \label{eq:bound_orders}
\begin{split}
	 &g(\rr,\kk) = f_{FD} \left( T,\mu, \En \left( \kk \right) \right)  \\
	 & - \frac{\tau \,e}{\hbar}  \cdot \nabla_{\kk} f_{FD}\left( T,\mu,\En\left(\kk\right)\right)  \cdot E  \\
	 & + \frac{\tau}{\hbar} \frac{\En-\mu}{T}   \nabla_{\kk} f_{FD}\left(T,\mu, \En\left(\kk\right)\right)  \cdot \nabla_{} T   \\
	 &  + \frac{\tau}{\hbar}   \nabla_{\kk} f_{FD}\left(T,\mu, \En\left(\kk\right)\right)\cdot \nabla_{} \mu  \\
	 & -  \frac{\tau^2}{\hbar^2}   \nabla_{\kk} \En \; \frac{\partial  }{\partial \mu}\nabla_{\kk} f_{FD}\left(T,\mu,\En\left(\kk\right)\right) 
	 (\nabla_{} \mu)^2 \\
	 & - \frac{\tau^2}{\hbar^2}   \nabla_{\kk} \En \; \nabla_{\kk} f_{FD}\left(T,\mu,\En\left(\kk\right)\right)  \frac{\partial^2 \mu}{\partial \rr^2} \\
	 &  - \frac{\tau^2}{\hbar^2}   \frac{\En-\mu}{T}  \nabla_{\kk} \En \; \nabla_{\kk} f_{FD}\left(T,\mu,\En\left(\kk\right)\right)
	 \frac{\partial^2 T}{\partial \rr^2} 
	 +  ...
\end{split}
\end{equation}
and it satisfies Eq.~\eqref{eq:Boltzmann_stat} but it does not satisfy the original BC. 
However, the difference  with respect to the full solution is exponentially small as soon as we move away from the boundary. 
Note, even though the terms $\delta g^{[\alpha]}[\rr,\kk, T, \mu]$ do not have an explicit  $\rr$-dependence far away from the boundary,  
they become position-dependent once we substitute for $T=T(\rr)$, $\mu=\mu(\rr)$, and $\EE=\EE(\rr)$
the functions obtained by solving self-consistently the continuity and Poisson equations (see Sec.\ref{sec:macrtransporteq}).

The first few terms in Eq.~\eqref{eq:bound_orders} coincide with the expressions for the Boltzmann distribution function  
in the presence of the  well known driving forces, i.e., we have reproduced, in a mathematically consistent way, 
the known textbook expressions\cite{Ziman_book,Ashcroft1976}. 
However, we also have the terms, like the last two, that have not been  reported before. 
These terms, together with similar, higher-order ones, are easy to overlook in the heuristic derivations 
that are often used to justify the first five terms. They only appear from the second order onwards and, therefore, 
usually give small corrections. However, the second order effects are not always negligible (for instance, the fifth term 
in Eq.~\eqref{eq:bound_orders} is sometimes very important).  
The last three terms can be important in inhomogeneous materials (say, multilayers) where the concentration 
and temperature vary rapidly across the sample.
In that case, all the terms of the same order should be treated on the same footing, i.e., 
one should not neglect the non-local forces proportional to the higher order derivatives of the temperature, chemical potential and electric field.

As shown in Appendix \ref{sec:hilbertexpansion}, the expansion obtained by neglecting the differential operator in Eq.~\eqref{eq:drift_nth_order} 
is the same as the one generated from the time-independent Boltzmann equation by the Hilbert 
expansion \cite{Hilbert.1917,Ukai.2006}, with the Knudsen number as the expansion parameter.  
(Knudsen number is given by the ratio of the mean free path, or the mean free time between collision,  
to some characteristic length (or time) of the system.) 
However, unlike the Hilbert expansion, our method retains its validity close to the boundary, 
provided we calculate the distribution function in each order from the differential drift-reaction equations \eqref{eq:drift_nth_order}.  In that case, the transport coefficients are not simply defined by local thermodynamic variables but have an explicit position dependence.

\section{Macroscopic transport equations } \label{sec:macrtransporteq}

In the previous section, we derived an approximate solution of the Boltzmann equation \eqref{eq:Boltzmann_stat} 
for arbitrary functions $T(\rr)$, $\mu(\rr)$ and $E(\rr)$. 
Substituting that solution in equations for the charge and energy conservation, Eqs.~\eqref{eq:number_cons} and \eqref{eq:energy_cons_0}, 
and using the Poisson equation \eqref{Eq:Poisson}, we can find the physical functions $T(\rr)$, $\mu(\rr)$ and $E(\rr)$.

Substituting the power series for $g(\rr,\kk)$ 
in Eqs.~\eqref{eq:charge_current} and \eqref{eq:energy_current}, 
yields the transport equations 
\begin{align} \label{eq:expanded_charge_current}
	\JJ_{}^{}=&\JJ^{[\nabla_{} T]}+\JJ^{[\nabla_{} \mu]}+\JJ^{[E]}+\JJ^{[E^2]}+\JJ^{[\nabla_{} E]}+...  \\
	\JJ_{\En}^{}=&\JJ_{\epsilon}^{[\nabla_{} T]}+\JJ_{\epsilon}^{[\nabla_{} \mu]}+\JJ_{\epsilon}^{[E]}+\JJ_{\epsilon}^{[E^2]}+\JJ_{\epsilon}^{[\nabla_{} E]}+...
\end{align}
where,   $\JJ^{[\alpha]}$ is  the charge current density due to the driving force $\xi_\alpha$, 
\begin{equation} \label{eq:generic_charge_current}
	\JJ^{[\alpha]}
	=
	e \left(\int \frac{\nabla_{\kk} \En }{\hbar} \delta g^{[\alpha]} \left(\rr,\kk \right) d^3k \right) \xi_\alpha 
	=  N_{\alpha}^J \xi_\alpha ~, 
\end{equation}
$\JJ_{\En}^{[\alpha]}$ is the corresponding energy current density, 
\begin{equation} \label{eq:generic_energy_current} 
	\JJ_{\En}^{[\alpha]}
	=
	\left( \int \En \frac{\nabla_{\kk} \En }{\hbar} \delta g^{[\alpha]} \left(\rr,\kk \right) d^3k  \right) \xi_\alpha 
	=  N_{\alpha}^\En \xi_\alpha ~, 
\end{equation}
and  $N_{\alpha}^J$ and $N_{\alpha }^\En$ are the transport coefficients associated with the force  $ \xi_\alpha$.  
Since the charge and energy conservation imply the continuity equations for the charge and energy currents, 
we can equivalently obtain $T(\rr)$, $\mu(\rr)$ and $E(\rr)$ by solving the continuity equations,  
$\nabla_{}\cdot \JJ(\rr)=0$ and $\nabla_{}\cdot \JJ_{\En}(\rr)=\JJ(\rr)\cdot\EE(\rr)$, together with 
the Poisson equation, and the transport equations, Eqs.~\eqref{eq:expanded_charge_current} and \eqref{eq:expanded_energy_current}. 

The above current densities and transport coefficients reproduce all the standard results for the response due to the known driving forces. 
For instance, the current which is first order in the electric field $\JJ^{[E]}=\sigma \EE$ has the conductivity  coefficient 
\begin{equation}
	\sigma= - \frac{\tau \,e^2}{\hbar^2}  \int \nabla_{\kk} \En\; \nabla_{\kk} f_{FD}\;d^3k  , \label{eq:conductsimple}
\end{equation}
which is the textbook result. The same agreement is found for the Seebeck coefficient $S$ and the thermal conductivity $\kappa$. 
We now report a few higher-order terms generated by the expansion of the distribution function given 
by Eq.~\eqref{eq:expansion_subst}. 
The current due to the second power of the electric field  is $\JJ^{[\EE^2]}=\sigma^{[\EE^2]}\EE^2$,  
where the conductivity coefficient, 
\begin{equation} \label{eq:charge_current_dmu2}
	\sigma^{[\EE^2]}= \frac{e^3\tau^2}{\hbar^3}\int \nabla_{\kk} \En \; \nabla_{\kk} \En  \;\nabla_{\kk} \En \; \frac{\partial^2  f_{FD}}{\partial \mu^2} d^3k  ~, 
\end{equation}
describes the Boltzmann expression for the non-linear response to an electric field.
The current driven by  $ \nabla_{}\EE(\rr) $  is $\JJ^{[\nabla_{}\EE]}=\sigma^{[\nabla_{}\EE]}\nabla_{}\EE$, 
with the conductivity coefficient 
\begin{equation} \label{eq:charge_current_d2mu}
	\sigma^{[\nabla_{}\EE]}=  \frac{e^2\tau^2}{\hbar^3} \int \nabla_{\kk} \En \;   \nabla_{\kk} \En \; \nabla_{\kk} \En \; \frac{\partial  f_{FD}}{\partial \mu}  d^3k ~, 
\end{equation}
while the current driven by  $ \partial^2 T/\partial \rr^2 $ is 
$\JJ^{[\frac{\partial^2 T}{\partial \rr^2}]}=\alpha^{[\frac{\partial^2 T}{\partial \rr^2}]}\partial^2 T/\partial \rr^2$, 
with the thermal conductivity coefficient 
\begin{equation} \label{eq:charge_current_d2T}
	\alpha^{[\frac{\partial^2 T}{\partial \rr^2}]}= \frac{e\tau^2}{\hbar^3} \int \frac{\En-\mu}{T} \nabla_{\kk} \En \;   \nabla_{\kk} \En \; \nabla_{\kk} \En \; \frac{\partial  f_{FD}}{\partial \mu}  d^3k.
\end{equation}
The other contributions to the currents, with the corresponding transport coefficients, can be calculated in the same way. 

The first few terms for the bulk current density read 
\begin{widetext}
\begin{equation}\label{eq:macroscopitransportcharge}
\JJ= \sigma(\EE +\frac{1}{e}\nabla_{}\mu) + \alpha \nabla_{} T + \sigma^{[\EE^2]} \EE^2+ \frac{\sigma^{[\EE^2]}}{e^2} \left(\nabla_{}\mu\right)^2 + \sigma^{[\nabla_{}\EE]}\nabla_{}\EE + \frac{\sigma^{[\nabla_{}\EE]}}{e} \frac{\partial^2 \mu}{\partial \rr^2} + .... ~, 
\end{equation} 
\end{widetext}
where the first term describes the response to a electro-chemical force, 
the second term describes to the Seebeck effect due to the thermal gradient, 
the third and the fourth term give the second-order response to the gradients of the electrical and chemical potentials,  
while the remaining terms describe the non-local response. 
A similar expression can be written for the heat current. 
Thus, our expansion supplements the well-known steady state macroscopic transport equations by additional terms 
which are due to the higher powers of the thermodynamic forces and their spatial derivatives. 
The  difference with respect to the usual textbook equations is the appearance of new, higher-order terms.

The current in Eq.~\eqref{eq:macroscopitransportcharge} does not have the 0-th order term, since it can be proven to vanish
because of the periodicity of the band structure and the fact that the integrand for the zeroth-order current is an exact differential. 
Let us also note that even-order contributions to the current densities are obtained by integrating 
the odd powers of the velocity. Such terms can be  finite  
if inversion and time reversal symmetry is broken, e.g.,
for a ferromagnet and a lattice without inversion symmetry. 
Otherwise  $ \En_\uparrow(-\kk) = \En_\downarrow(\kk)$ so that there is no
charge current, but possibly a spin current if inversion symmetry is broken.

\section{Example: Depletion region in semiconductors}  \label{sec:examplesemicond}

\begin{figure*}[t]
 \includegraphics[width=0.8\textwidth]{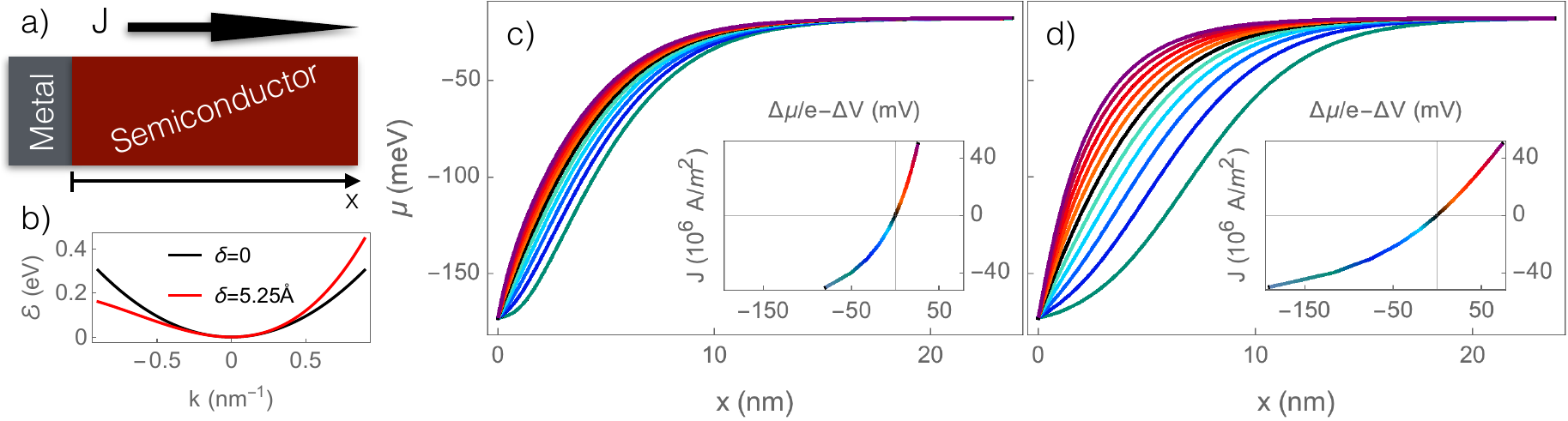}
 \caption{[Colour online] 
Panel a) Geometry of the metal-semiconductor junction and the direction of positive current. The interface is at $x=0$. 
Panel b) Modification of the bottom of the band structure as in Eq.~\ref{eq:asymmbandstructure}.
Panel c) Position-dependent chemical potential calculated for different currents at $T=300K$ for the band structure with a finite $\delta$, but
using the textbook transport equations, i.e.~setting $\sigma^{\scriptscriptstyle [\nabla_{} \EE]}=0$ in the complete set of equations. 
The results depend on the value of the current running through the device (see the inset, which also provides the color code for the currents). 
Panel d) Same as panel c) but calculated with the correct transport equations, including the new term $\sigma^{\scriptscriptstyle [\nabla_{} \EE]}$. 
 The insets in c) and d) provide the current vs voltage characteristic of the semiconductor layer.
}\label{fig:metalsemi}
\end{figure*}

 To show the impact of the new terms on the behavior of real devices we  now consider an example involving 
the terms proportional to $\partial^2 \mu/\partial \rr^2$ and $\nabla_{} \EE$. 
These terms are proportional to $\tau^2$, so $\sigma^{[\nabla_{}\EE]}$ is usually small, 
but if the scattering life time is  long or the derivatives of the potential are large, 
their contribution can be important. 
Taking the case that is familiar to most readers, we examine the width and the shape of the depletion region in a metal-semicondutor (M-S) junction 
shown in Fig.~\ref{fig:metalsemi}.a.  
We are only interested in the qualitative features due to the new terms, revealed by our treatment,  so we neglect several effects 
that are relevant for real junctions, like finite jumps in temperature and chemical potential or the formation of defects. 

Let us, first,  recapitulate the textbook approach to isothermal M-S junctions, ignoring the heat transport \cite{Sze}. 
The junction is described by two equations: 
\begin{align}
	&J= \sigma(\mu(x))\left( \EE(x)-\frac{\mu'(x)}{e}\right) = \mbox{const.}   \label{eq:1Dtextbookcurrent} \\
	&\epsilon_0 \EE' (x)=  \rho(\mu(x))
\end{align}
where the first one is the trasnsport  equation ~\eqref{eq:macroscopitransportcharge} and  the second one is the Poisson equation (Eq.~\eqref{Eq:Poisson}). 
As shown below,  the conductivity $\sigma$ and the charge $\rho$ depend  explicitly on the chemical potential, 
so that the solution of these coupled equations yields, for a constant current density $J$, the spatial profile of the electric field and the chemical potential. 
The conductivity is calculated from Eq.~(\ref{eq:conductsimple}) and to obtain an explicit expression for $\sigma(x)$ and $\rho(x)$  
we consider an n-type semiconductor with a parabolic band  $\En(k)=\hbar^2 k^2/2m$. 
We also assume that the chemical potential is below the bottom of the conduction band and approximate the Fermi-Dirac distribution 
by $f_{FD}\approx\exp[(\mu-\En)/k_B T]$. The conductivity and the charge $\rho$ (see Eq.~\eqref{eq:elnumber}) assume the simple form
\begin{align}
	\sigma &= \sigma_0 e^{\frac{\mu}{k_B T}} \label{eq:condsemicond}\\
	\rho &= \rho_D -\rho_0 e^{\frac{\mu}{k_B T}}  \label{eq:chargesemicond}
\end{align}
where $\sigma_0$ and $\rho_0$ are given by the expressions, 
\begin{align}
	& \rho_0 = \frac{\sqrt{2 \pi k_B T m}}{ \hbar},\\
	& \sigma_0 =  \frac{\tau e^2 }{\hbar} \sqrt{\frac{2\pi k_B T}{ m}} ,
\end{align}
and $\rho_D$ is the dopant charge density.
Thus, we obtain the textbook equations for the charge transport in semiconductors, 
\begin{align}
	&J= \sigma_0 e^{\frac{\mu(x)}{k_B T}}\left( \EE(x)-\frac{\mu'(x)}{e}\right)   ,\label{eq:currentsemicond}\\
	&\epsilon_0 \EE' (x)=  \rho_D -\rho_0 e^{\frac{\mu(x)}{k_B T}} \label{eq:Poisssemicond}
\end{align}
which have to be solved for the appropriate boundary conditions. 
For a complete description, we also need similar equations for the metal part of the junction (with the appropriate 
expressions for the conductivity and the charge) and, then, we have to link the two regions. 
The requirement is that the chemical potential and the electric field (or the electric displacement field, when the 
semiconductor and the metal have different dielectric constants) are continuous across the interface  
(again, we neglect the surface discontinuities in the potentials). 
However, if the metal is highly conductive  and has a high density of states at the Fermi energy, 
it is sufficient to solve Eqs.~\eqref{eq:currentsemicond} and \eqref{eq:Poisssemicond} 
with the boundary conditions $\mu(0)=\En_{F,m}$ and $\mu(+\infty)=k_B T \log (\rho_D/\rho_0)$.  
That is, we require that the chemical potential at the interface is set by the Fermi level $\En_{F,m}$  of the metal 
and that there is no net charge far away from the interface.

We consider a low symmetry material with the dispersion (close to the band minimum) given by 
\begin{equation} \label{eq:asymmbandstructure}
	\En(\kk)=\frac{\hbar^2 (k^2 +\delta \, k_x^3)}{2m} 
\end{equation}
where $k$ is the magnitude of the crystal momentum $\kk$, $k_x$ its the $x$ component, and $\delta$ measures the asymmetry. 
The modification of band structure is depicted in Fig.~\ref{fig:metalsemi}b. 
The difference with respect to the parabolic dispersion is that  $\En(\kk)$ now has a finite third derivative at the minimum.
As an example, we take a semiconductor similar to silicon ($m=0.1 m_e$, where $m_e$ is the electron mass, 
and $\tau=10$ fs) but with an asymmetry $\delta=5.25$\AA, doped with $3.6*10^{22}$ carriers$/m^3$,  
and attach it to a metal with Fermi level at $-155 ~ meV$ below the semiconductor conduction band.

The solution of Eqs.~\eqref{eq:currentsemicond} and \eqref{eq:Poisssemicond} provides the spatial profile of the 
electric field and the chemical potential. The results are shown in Fig.~\ref{fig:metalsemi}c for various operating conditions. 
The black line represents an open circuit (no current flowing) and shows the formation of the depletion region. 
As the current increases, the depletion region either expands, for negative currents, or shrinks, for positive ones. 
The inset shows a typical current-voltage characteristic of a metal-semiconductor junction, i.e.~it shows the dependence 
of the electro-chemical potential in the semiconductor, $\mu(+\infty)/e-V(+\infty)  - (\mu(0))/e-V(0))$, 
on the current running through the junction.

We now compare this textbook solution with the case when the response of a semiconductor in the presence of 
additional driving forces  $\partial^2 \mu/\partial \rr^2$ and $\nabla_{} \EE$. 
It is easy to prove that
 \begin{equation}
	\sigma^{\scriptscriptstyle [\nabla_{} \EE]} = \sigma_0^{\scriptscriptstyle [\nabla_{} \EE]} e^{\frac{\mu}{k_B T}}\; ,
	\label{eq:cond2semicond}
\end{equation}
where $\sigma_0^{\scriptscriptstyle [\nabla_{} \EE]}$ is a constant, 
and that $\sigma^{\scriptscriptstyle [\nabla_{} \EE]}$ vanishes for centrosymmetric materials ($\delta=0$). 
That is,  the shape  of the conductivity and charge is still given by Eqs.~\eqref{eq:condsemicond} and \eqref{eq:chargesemicond} 
but the proportionality constants are now more complex.
The  macroscopic transport equation, including the new terms, reads 
\begin{align}
	&J = \sigma_0 e^{\frac{\mu}{k_B T}}\left( \EE-\frac{\mu'}{e}\right) + \sigma^{\scriptscriptstyle [\nabla_{} \EE]}_0e^{\frac{\mu}{k_B T}}\left( \EE'-\frac{\mu''}{e}\right) \;, 
	\label{eq:currentdevice}
\end{align}
while the Poisson equation is unchanged.
As before, we also have to consider  the transport equations in the metal and ensure that the electric field 
(or electric displacement field if the semiconductor and the metal have different dielectric constants), 
the chemical potential, and its first derivative are continuous across the interface. 

In the absence of the gradient of the  electro-chemical potential we can set $\sigma^{\scriptscriptstyle [\nabla \EE]}_0=0$ 
and  reduce Eq. \eqref{eq:currentdevice} to Eq.~\eqref{eq:currentsemicond}. 
When $\sigma^{\scriptscriptstyle [\nabla_{} \EE]}_0\neq 0$, the order of the differential equation rises 
and the boundary condition for the derivative is needed. We require that the first derivative has the same 
value as in the case $\sigma^{\scriptscriptstyle [\nabla_{} \EE]}_0=0$ and report the  solution of Eqs.~\eqref{eq:currentdevice} 
and \eqref{eq:Poisssemicond}  in Fig.~\ref{fig:metalsemi}c.  The comparison of Figs.~\ref{fig:metalsemi}c and \ref{fig:metalsemi}d  shows that in the absence of the  current, 
the textbook equations and the complete transport equations lead to  the same result (the black curves in Fig.~\ref{fig:metalsemi}c 
and \ref{fig:metalsemi}d are the same), as can be deduced from Eq.~\eqref{eq:currentdevice} for $J=0$. 
However, for $J\neq 0$, the dimension and shape of the depletion region are modified by the new terms, 
i.e., the linear response theory differs considerably from the full description of the device. 
For negative currents, the  depletion region given by the non-linear description is wider than the textbook one, 
while for positive currents, the depletion region is reduced. 

The above concepts can have a straightforward experimental verification in the case of a perfectly symmetric 
metal-semiconductor-metal device, when the semiconductor has both  inversion and time reversal symmetries broken. 
The standard transport equations predict that when the current is flowing in positive direction, the voltage drop is exactly 
the opposite to the one when the direction of the current is reversed. 
Thus,  the standard equations predict that the two running conditions  are perfectly symmetric. 
However, if the terms proportional to the derivative of the electric field and the second derivative 
of the chemical potential are taken into account we expect the asymmetric behaviour, 
as can be verified by applying the transformation $x \rightarrow -x$ to Eq.~\eqref{eq:currentsemicond} and \eqref{eq:currentdevice}.

We provided just one example of possible effects due to the new terms obtained by the expansion of the Boltzmann distribution function. 
But the range of applicability and the relevance of that expansion is much wider and we expect that other terms will also become important.

\section{Discussion and conclusions}  \label{sec:conclusion}


In summary,  our starting point is the stationary Boltzmann equation, in its most general form, subject to specific microscopic boundary conditions. 
To obtain the solution, we carefully define the relaxation time approximation (RTA) and relate, for a given spatial profiles of temperature $T(\rr)$, 
chemical potential $\mu(\rr)$, and electric field $\EE(\rr)$, the Boltzmann distribution function  to the relaxation time.  
 We then show that to obtain the physically meaningful results  $\EE(\rr)$ has to satisfy the Poisson equation, 
while $T(\rr)$ and $\mu(\rr)$  have to satisfy the charge and energy conservations. 
Thus, together with the the Boltzmann equation, we now have to self-consistently solve three additional equations. 
The initial problem for  $g(\rr,\kk)$ has, apparently, been turned into a more complicated one, 
where $g(\rr,\kk)$ is a functional defined on the functions $T(\rr)$, $\mu(\rr)$, and $\EE(\rr)$. 

We now focus on the strategy for solving the Boltzmann equation within the RTA, by assuming that $T(\rr)$, $\mu(\rr)$, and $\EE(\rr)$, are analytic functions. 
Representing these functions by their respective Taylor series\cite{functional}, we can write $g(\rr,\kk)$ as a function of infinitely many variables, 
\begin{align} \label{eq:op}\nonumber
	g(\rr,\kk)=\tilde{g} \big(&\rr,\kk,  T(\rr),\mu(\rr), \EE(\rr) , \nabla T(\rr),\nabla \mu(\rr),\\
	& \nabla \EE(\rr), \frac{\partial^2 T(\rr)}{\partial \rr^2},\frac{\partial^2 \mu(\rr)}{\partial \rr^2},\frac{\partial^2 \EE(\rr)}{\partial \rr^2}, ..., BC\big) .\nonumber
\end{align}
The general solution of the Boltzmann equation is obtained by expanding the distribution function 
with respect to all its variables, except the first five, $\rr,\kk, V(\rr), T(\rr)$, and $\mu(\rr)$, and writing $g(\rr,\kk)$ as a multivariable power series. 
Since the expansion variables are completely arbitrary, substituting $g(\rr,\kk)$ in the Boltzmann equation 
yields an infinite number of coupled differential equations for the expansion coefficients. 
Integrating these equations for a particular set of microscopic boundary conditions  yields $g(\rr,\kk)$  as a power series in terms of the expansion variables 
$\EE(\rr) , \nabla_{} T(\rr),\nabla_{} \mu(\rr),  \nabla_{}^2 V(\rr),  \nabla_{}^2 T(\rr), \nabla_{}^2 \mu(\rr) \ldots$. 
Substituting $g(\rr,\kk)$ in the expressions for the charge and energy current densities we obtain an expansion of $\JJ(\rr)$ and $\JJ_\En(\rr)$ 
in  terms of their respective driving forces.
The coefficients of the driving forces define the generalized transport coefficients. 
The physically relevant functions $\JJ(\rr)$, $\JJ_\En(\rr)$, $\mu(\rr)$, $T(\rr)$, and $\EE(\rr)$ are obtained at every macroscopic point of the sample,   
by solving self-consistently the transport equations, the charge and energy continuity equations, and the Poisson equation. 
 We also show, under which conditions the surface effects can be neglected and 
the distribution function of a macroscopic sample  assumes a simple, textbook form.

The above procedure elucidates the commonly used derivation of transport equations and exposes various approximations employed in that derivation. 
In addition, 
it reveals new contributions to the response functions which  are proportional to the higher powers of the  forces
and their higher-order derivatives. The ensuing corrections to the charge and energy currents are usually small, which explains 
the success  of the phenomenological transport theory in Eqs.~\eqref{eq:phentranspincomplcharge} and \eqref{eq:phentranspincomplenergy}.
However, in certain situations, the new terms lead to qualitatively new phenomena. 
For example, they can become important for heterogeneous devices, for materials 
in which the transport properties are strongly temperature- and potential-dependent, 
for systems driven out of equilibrium by large thermodynamic forces 
(e.g. large temperature differences), or when the thermodynamic potentials vary strongly over small distances. 
Unlike the second order  response to the thermodynamic forces, like the one due to $\EE^2$, $(\nabla \mu)^2$ or $(\nabla T)^2$, 
the response to the spatial derivatives of these forces, like $\nabla_{} \EE$, $\nabla_{}^2 \mu$,  or $\nabla_{}^2 T$, has not been discussed before. 
Since the magnitude of the new higher-order terms is comparable to the already known ones, all the terms of the same order should be treated on the same footing. 
In other words, a consistent semi-classical description of transport phenomena should not just consider the higher powers of the thermodynamic forces but 
 should also take into account the driving forces which are proportional to the higher order derivatives of temperature, chemical potential and electric field. 

The expansion of the distribution function described in this paper respects the microscopic boundary conditions  of the Boltzmann equation. 
Thus, it can be used to treat, on the same footing, not just the higher-order and non-local effects in the bulk but also the multitude of surface effects.
Our expansion in terms of the driving forces provides a substantial improvements over the Hilbert expansion or similar expansions 
of the solution of the Boltzmann equation which do not take into account the microscopic boundary conditions, 
and therefore yield the solution which is valid only  in the bulk.

\begin{acknowledgments}
The authors wish to thank Jan Tomczak and Michael Wais for fruitful discussions, and  acknowledge financial support by the European Research Council/ERC through grant agreement n.\ 306447 and by the Austrian Science Fund (FWF) through  SFB ViCoM F41 and Lise Meitner position M1925-N28. V.Z acknowledges the support by the Ministry of Science of Croatia under the bilateral agreement with the USA on the scientific and technological cooperation, Project No. 1/2016. 
\end{acknowledgments}

\appendix
\section{Continuity equations} \label{sec:continuityEquations}

The particle continuity equation is obtained by integrating Eq.~\eqref{eq:Boltzmann} over  the whole $k$-space. The first term of Eq.~\eqref{eq:Boltzmann} becomes the time evolution of the local total number of particles, defined as
\begin{equation}
	n(t,\rr)=\int g \left(t, \rr,\kk \right) d^3k. \label{eq:elnumber}
\end{equation}
In the second term of Eq.~\eqref{eq:Boltzmann} the divergence with respect to spacial coordinates can be brought out of the $k$-space integral, leading to the spatial divergence of the particle current written as 
\begin{equation}
	\JJ(t,\rr) = e \int \frac{\nabla_{\kk} \En }{\hbar} \, g\left(t, \rr,\kk \right)  d^3k.
\end{equation}
The third term of Eq.~\eqref{eq:Boltzmann} can be proven to integrate to zero. The part multiplying the electric field $\EE=\nabla_{} V + \partial \AAA/\partial t$ vanishes due to a corollary of the divergence theorem in the $k$-space and the periodicity of all the involved functions in $k$-space. The part multiplying the magnetic field $\BB= \nabla_{} \times \AAA$ requires the use of the identity $\nabla_{\kk} \cdot ( \nabla_{\kk} \En \times \BB) =0$ and then the same considerations  above. 

The integral of the scattering term on the right hand side of Eq.~\eqref{eq:Boltzmann} depends on its precise expression. We assume the RTA in Eq.~\eqref{eq:relaatimeapprox0}. As already mentioned in section \ref{sec:TransportEquations}, the value of the integral will depend on the local temperature $T(\rr)$ and chemical potential $\mu(\rr)$, as well as the local electron distribution $g\left(t, \rr,\kk \right)$ and can be, in general, different from zero. This would imply that some particles are either destroyed or created during the scattering. It is indeed to prevent this unphysical effect that we imposed the constraint in Eq.~\eqref{eq:number_cons}, for the RTA to make sense. Using Eq.~\eqref{eq:number_cons}, the conservation equation reduces to
\begin{equation} \label{eq:particel_continuity}
	\frac{\partial }{\partial t} \int g \left(t, \rr,\kk \right) d^3k +\nabla_{}\cdot  \int \frac{\nabla_{\kk} \En }{\hbar} g \left(t, \rr,\kk \right) d^3k =0.
\end{equation}
which states that the variation in the local number of particles $n$, has to be equal to the divergence of the particle current density $\JJ(\rr)$.

Similarly the energy continuity equation can be obtained by multiplying Eq.~\eqref{eq:Boltzmann} by the particle energy $\En(\kk)$ and integrating over  the whole $k$-space. The first term will give the change in the 
energy density $\epsilon(t,\rr)=\int \En(\kk) \,g \left(t, \rr,\kk \right) d^3k$. The second yields  the divergence of the energy current density $\JJ_{\epsilon}(t,\rr) =  \int  \En \, g \,\nabla_{\kk} \En /\hbar \, d^3k$. The third term in this case does not vanish but is the work made by the electrical field on the system $W(t,\rr)=e/\hbar \int \En\, \EE \cdot \nabla_{\kk} g \,d^3k $ leading to the Joule heating. Again we use the RTA for the scattering term. Its integral is now constrained by Eq.~\eqref{eq:energy_cons_0} (eventually with the effect of phonons included as explained in the text below Eq.~\eqref{eq:energy_cons_0}).We therefore obtain:
\begin{equation} \label{eq:energy_continuity}
\begin{split}
	&\frac{\partial }{\partial t} \int \En \, g \left(t, \rr,\kk \right) d^3k +\nabla_{}\cdot  \int \En \frac{\nabla_{\kk} \En }{\hbar} g \left(t, \rr,\kk \right) d^3k \\
	&\;\;\;\;\;\;\;\;\;+\frac{e}{\hbar}  \int \nabla_{\kk} \En \cdot  \nabla_{} V   g \,d^3k =    \frac{\Delta \epsilon_{e-ph} }{\tau} . 
\end{split}
\end{equation}
where we have integrated by parts the Joule heating term and used again the periodicity of all the involved functions in $k$-space. The equation implies that any change in the local total energy $\epsilon$, is caused by the divergence of the energy current $\JJ_{\epsilon}$ and the work $W$ done on the charged particles by the electric field and the energy dissipated into phonons $\Delta \epsilon_{e-ph}$. \\

\section{Comparison with Hilbert expansion} \label{sec:hilbertexpansion}
Some of the results derived in this paper can also be obtained from the Hilbert expansion for the Boltzmann equation in the relaxation time approximation. 
However, as shown below, our approach overcomes one fundamental limitation of the Hilbert expansion, which critically limits its range of applicability 
(as well as its use in the longstanding mathematical problem of the proof of the existence of the solution of the Boltzmann equation in presence of the boundary conditions).

By adapting the Hilbert expansion to the present case, we look for the solution of Eq.~\ref{eq:Boltzmann_stat} in the form:
\begin{equation}
                        \label{eq:hilberexpansion}
	g\left(\rr,\kk \right) = \sum_{i=0}^{\infty} \tau^i g^{[i]}_H\left(\rr,\kk \right)
\end{equation}
where here the relaxation time $\tau$ plays the role of the Knudsen parameter. Substituting the above series in the Boltzmann equation and 
collecting all the terms of the same order in $\tau$ gives the result 
\begin{equation} \label{eq:orderhilbert}
\begin{split}
	g^{[0]}_H\left(\rr,\kk \right)&= f_{FD} \left( T\left(\rr\right), \mu\left(\rr\right), \En \left( \kk \right) \right) \\
	g^{[1]}_H\left(\rr,\kk \right)&=- \frac{1}{\hbar}   \nabla_{\kk} \En  \cdot \nabla_{\rr} g^{[0]}_H +\frac{e}{\hbar}  \EE \cdot \nabla_{\kk} g^{[0]}_H \\
	g^{[2]}_H\left(\rr,\kk \right)&=- \frac{1}{\hbar}   \nabla_{\kk} \En  \cdot \nabla_{\rr} g^{[1]}_H +\frac{e}{\hbar}  \EE \cdot \nabla_{\kk} g^{[1]}_H \\
	...&=...
\end{split}
\end{equation}
Substituting successively the lower-order corrections into the higher-order ones, we find that the distribution function defined by 
Eq.~\ref{eq:orderhilbert} is equivalent to the one given by Eq.~\ref{eq:bound_orders}. 
Note, the expansion in Eq.\eqref{eq:hilberexpansion}  does not take into account the boundary conditions, so that it provides just one of the 
infinitely many solutions of the Boltzmann equation.  
In fact, the result given by Eqs.~\eqref{eq:orderhilbert}  or \eqref{eq:bound_orders} corresponds 
to a specific choice of the boundary conditions, as discussed in Sec.\ref{sec:macrodevices}. 
Hence, the solution obtained by the Hilbert expansion does \emph{not}, in general, satisfy the imposed boundary conditions
 
This shows the most important difference between the approach taken in this work and the one taken by Hilbert expansion. 
Each term in the expansion defined by Eq.~\ref{eq:expansion_subst} of this work satisfies one of the differential equations given by Eqs.~\eqref{eq:zero_order}-\eqref{eq:drift_react_d2mu} and the sum of all these terms provides a particular solution of the Boltzmann equation 
that  complies with the imposed boundary conditions. 
Thus, the expansion in terms of the driving forces presented in this work is much more powerful than the Hilbert expansion.

\bibliography{Biblio.bib}

\end{document}